\documentclass[aps,prx,bibliography,superscriptaddress,,twocolumn]{revtex4}
\usepackage[utf8]{inputenc}
\usepackage{amsmath}
\usepackage{physics}
\usepackage{mathtools}
\usepackage{amsfonts}
\usepackage{amssymb}
\usepackage{braket}
\usepackage{graphicx}
\usepackage{subfigure}
\usepackage{textcomp}
\usepackage{color}
\usepackage[dvipsnames]{xcolor}
\usepackage{mathrsfs}
\usepackage{natbib}
\usepackage{makecell}
\usepackage{soul}

\begin{document}
\title{N\'eel vector controlled exceptional contours in $p$-wave magnet-ferromagnet junctions}
\author{Md Afsar Reja}
\email{afsarmd@iisc.ac.in}
\affiliation{Solid State and Structural Chemistry Unit, Indian Institute of Science, Bangalore 560012, India}
\author{Awadhesh Narayan}
\email{awadhesh@iisc.ac.in}
\affiliation{Solid State and Structural Chemistry Unit, Indian Institute of Science, Bangalore 560012, India}
\date{\today}

\begin{abstract}
Non-Hermitian systems can host exceptional degeneracies where not only the eigenvalues, but also the corresponding eigenvectors coalesce. Recently, $p$-wave magnets have been introduced, which are characterized by their unusual odd parity. In this work, we propose the emergence of non-Hermitian degeneracies at the interface of $p$-wave magnets and ferromagnets. We demonstrate that this setup offers a remarkable tunability allowing realization of exceptional lines and rings, which can be controlled via the orientation of the $p$-wave N\'eel vector. We present the origin of these exceptional contours based on symmetry, and characterize them using phase rigidity. Our works puts forward a versatile platform to realize controllable non-Hermitian degeneracies at odd parity magnetic interfaces.
\end{abstract}

\maketitle


\noindent \textit{\textbf{Introduction--}} Non-Hermitian (NH) systems exhibit a wide range of novel physical phenomena, such as exceptional points (EPs), the non-Hermitian skin effect, and unconventional topological phases -- features that have no direct analogue in Hermitian systems~\cite{moiseyev2011non,bender2007making,ashida2020non,bergholtz2021exceptional,kawabata2019symmetry,banerjee2023non,ding2022non,el2018non,okuma2023non,meng2024exceptional}. Exceptional degeneracy is a unique type of degeneracy exclusive to NH systems, where both eigenvalues and eigenvectors merge, unlike Hermitian degeneracies, where only eigenvalues coincide. These degenerate points are known as EPs~\cite{heiss2012physics,kato2013perturbation}, and their extended structures -- exceptional rings (ERs) -- have also been actively studied~\cite{yoshida2019symmetry,wang2019non,bergholtz2021exceptional, liu2021higher,xu2017weyl}. ERs have also been realized in a few experimental platforms, including photonic~\cite{cerjan2019experimental,zhen2015spawning} and thermal diffusive systems~\cite{xu2022observation}. EPs have attracted increasing interest because of their theoretically rich prospects~\cite{ding2022non,banerjee2023tropical} and promising applications in various areas of physics, including photonics, acoustics, sensing, and electric
circuits~\cite{chen2017exceptional,hodaei2017enhanced, parto2020non,miri2019exceptional,stehmann2004observation,choi2018observation,zhu2018simultaneous,shi2016accessing,zhu2018simultaneous}. They have also been theoretically explored at various material interfaces~\cite{bergholtz2019non,cayao2022exceptional,cayao2023exceptional}.

Altermagnets (AMs)~\cite{vsmejkal2022emerging,vsmejkal2022beyond} have recently garnered considerable interest both in theoretical and experimental studies because of their unique combination of features arising from both ferromagnets (FM) and antiferromagnets. \textcolor{black}{AMs are characterized by a distinct form of magnetic ordering. In conventional antiferromagnets, the opposite spin sublattices are connected via translational or inversion symmetry operations. In contrast, in AMs, the opposite spin sublattices are related through non-centrosymmetric operations such as mirror reflections or rotational symmetries.}
Remarkably, AMs display spin-split electronic bands similar to those of FMs but maintain a net zero magnetization like antiferromagnets~\cite{vsmejkal2022beyond,litvin1974spin,brinkman1966theory}. These distinct properties make them promising candidates for spintronics applications. A diverse class of materials has been identified as AMs, supported by both theoretical predictions and experimental confirmations~\cite{vsmejkal2022beyond,fedchenko2024observation,vsmejkal2022emerging,yuan2020giant,vsmejkal2022emerging,vsmejkal2022beyond,krempasky2024altermagnetic,zhu2024observation,devaraj2024interplay}. Junctions of AMs with different materials have revealed intriguing transport properties~\cite{sun2023spin,das2023transport,lyu2024orientation,papaj2023andreev,ouassou2023dc,beenakker2023phase,niu2024electrically,giil2024superconductor,cheng2024field,das2024crossed,sukhachov2024thermoelectric,niu2024orientation,lu2024varphi,banerjee2024altermagnetic,nagae2024spin,zhang2024finite,chen2024electrical,lin2025coulomb,sun2025tunable}. In NH context, the emergence of EPs at AM–FM junctions has been very recently predicted~\cite{reja2024emergence,dash2025fingerprint}.

\begin{figure}[b]
\centering
\includegraphics[width=0.32\textwidth]{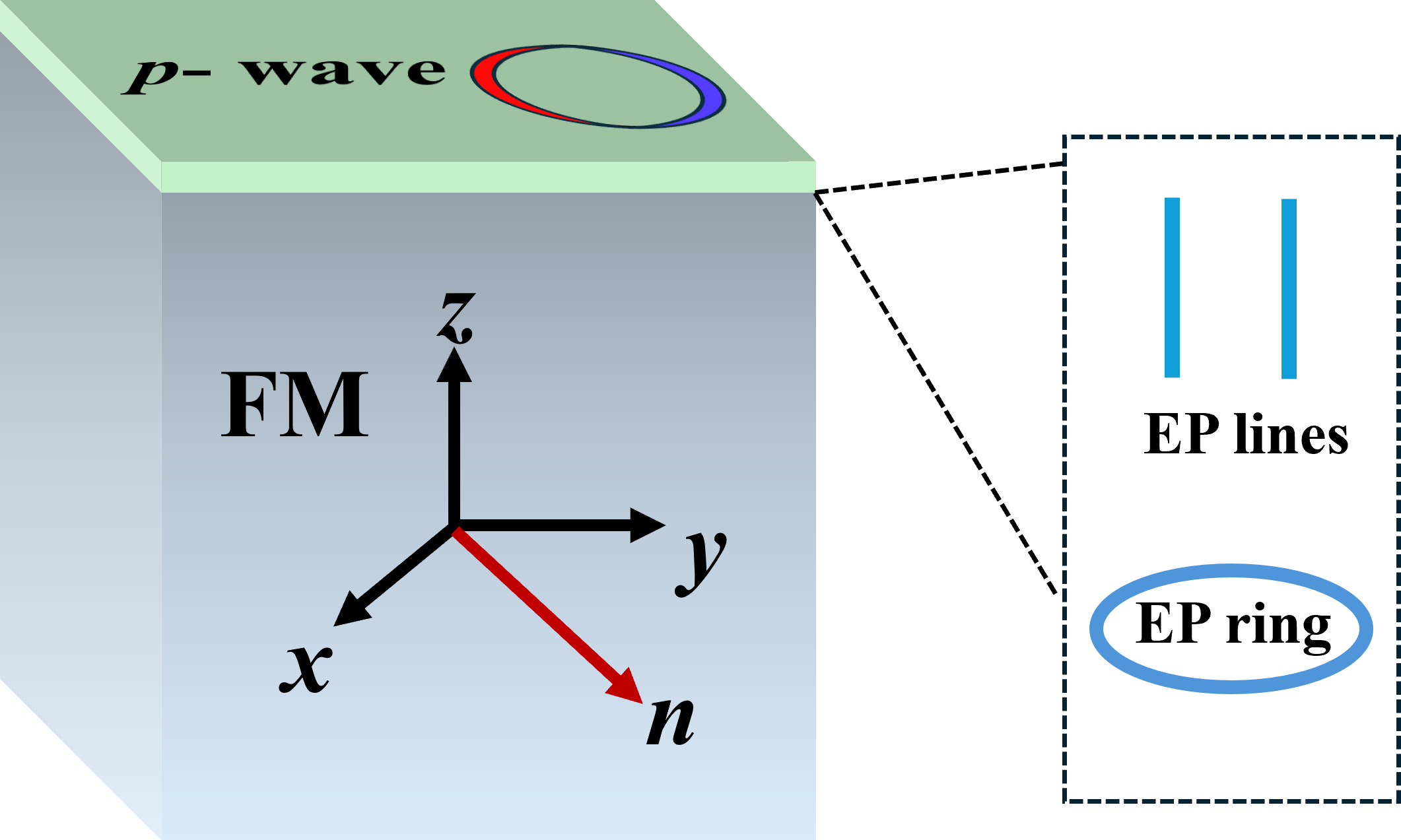}
\caption{\textbf{Proposed setup with $p$-wave magnet-FM junction.} Illustration of the $p$-wave magnet-FM junction at $z=0$, with FM region extending for $z<0$. The Fermi surface with red and blue color shaded regions shows the odd parity behavior of $p$-wave magnets. We propose the appearance of symmetry protected exceptional rings and exceptional lines at such junctions, as shown schematically in the dotted box.  \textcolor{black}{To occur such degeneracy the N'eel vector $\mathbf{n}$ (red arrow) must lie within the $x$–$y$ plane.}}
\label{Fig_junction}
\end{figure}

So far, all known AMs were categorized as even parity ($d$-, $g$-, or $i$-wave), based on the nodal surfaces of Fermi surfaces belonging to opposite spin sublattices~\cite{vsmejkal2022emerging,vsmejkal2022beyond}. However, a new class of AMs with odd parity -- termed $p$-wave magnets -- has very recently been identified~\cite{hellenes2023p,brekke2024minimal}. These $p$-wave magnets are distinct in terms of symmetry -- they break spatial inversion (parity) symmetry while preserving time-reversal symmetry (TRS). This is in contrast to conventional even-parity AMs, which break TRS. In addition to material realization~\cite{hellenes2023p,yamada2025gapping}, a number of interesting properties of $p$-wave magnets are starting to be uncovered~\cite{brekke2024minimal,ezawa2024purely,hedayati2025transverse,soori2025crossed,fukaya2025superconducting,hayami2019momentum}.

In this work, we propose the appearance of exceptional contours -- rings and lines -- arising at the junction composed of $p$-wave magnets and FMs. We find that the type of contours and their positions and shapes can be tuned via the orientation of the N\'eel vector. We further analyze the origin of the exceptional contours based on symmetry and characterize them using phase rigidity. Our work introduces a promising avenue for exploring non-Hermitian exceptional features in a new platform comprising of novel magnetic systems. \\


\noindent \textit{\textbf{Setup of AM-FM junctions--}} We couple a two-dimensional $p$-wave magnet to a semi-infinite FM lead as shown in Fig.~\ref {Fig_junction}. Here the interface lies at $z=0$, with the FM lead extending for $z<0$. The junction is modeled as an open quantum system, described by the following Hamiltonian,

\begin{equation}
\Tilde{H} = H_p + \Sigma_L.
\label{eq_H_NH}
\end{equation}

Here $H_\text{p}$ is the Hamiltonian of the two dimensional $p$-wave magnet (discussed later), and $\Sigma_L$ represents the self-energy arising from the semi-infinite FM lead. Under the wide-band approximation, the self-energy term does not depend on momentum or frequency and can be computed analytically as~\cite{ryndyk2009green,datta1997electronic,bergholtz2019non,cayao2022exceptional},

\begin{equation}
\Sigma_L=-i\Gamma\sigma_0 -i\gamma\sigma_z,
\end{equation}

where $\Gamma = \frac{\Gamma_{+} + \Gamma_{-}}{2}$, $\gamma = \frac{\Gamma_{+} - \Gamma_{-}}{2}$, and $\Gamma_{\pm} = \pi |t'|^2\rho_{\pm}^L$. Here $\rho_{\pm}^L = \frac{1}{t'\pi} \sqrt{1 - (\frac{\mu_L \pm m}{2t_z})^2}$ The quantities $\rho_{\pm}^L$ represent the surface density of states of the lead for the spin-up and spin-down channels, respectively. The parameter $t'$ denotes the hopping amplitude between the lead and the $p-$wave magnet. In this framework, $\sigma_x$, $\sigma_y$, and $\sigma_z$ denote the Pauli matrices, and $\sigma_0$ represents the $2 \times 2$ identity matrix. The parameter $t_z$ corresponds to the hopping amplitude along the $z$-direction within the lead. The quantity $\mu_L$ is the chemical potential of the lead, while $m$ characterizes the intrinsic magnetization of the FM lead. The coupling to the lead introduces an effective NH character into the effective Hamiltonian of the junction through the imaginary part of the self-energy. We next see how this results in exceptional physics at our proposed junction. \\


\noindent \textit{\textbf{Non-Hermitian degeneracies at $p$-wave magnet-FM junctions--}} We consider the above setup with a two-dimensional $p$-wave magnet with the Hamiltonian given by~\cite{ezawa2025out,ezawa2024purely,yamada2025gapping},

\begin{equation}
\begin{aligned}
     H_p=t(k_x^2+k_y^2)\sigma_0+ \lambda(-k_y\sigma_x+k_x\sigma_y)+ J(\mathbf{n}\cdot \boldsymbol{\sigma})k_x.
     \end{aligned}
\label{eq_pH}
\end{equation}

Here $t$ is the hopping amplitude, $\lambda$ is the strength of the Rashba term, and $\mathbf{n}=(\sin\theta\cos\phi,\sin\theta\sin\phi,\cos\theta)$ is the N\'eel vector with magnitude $J$. The first term describes free electrons, the second term corresponds to Rashba spin-orbit coupling, and the third term represents the $p$-wave characteristic contribution, which is linear in momentum. For the $p$-wave-FM junction, the effective NH Hamiltonian then becomes, 

\begin{equation}
 \begin{aligned}
        \Tilde{H} =& H_p + \Sigma_L\\
              &= t(k_x^2 + k_y^2) \sigma_0 + \lambda(-k_y\sigma_x + k_x\sigma_y)+  J(\mathbf{n}\cdot \boldsymbol{\sigma})k_x + \Sigma_L.   
\end{aligned}
\label{eq_d_ham}
\end{equation}

\begin{figure}
\centering
\includegraphics[width=0.48\textwidth]{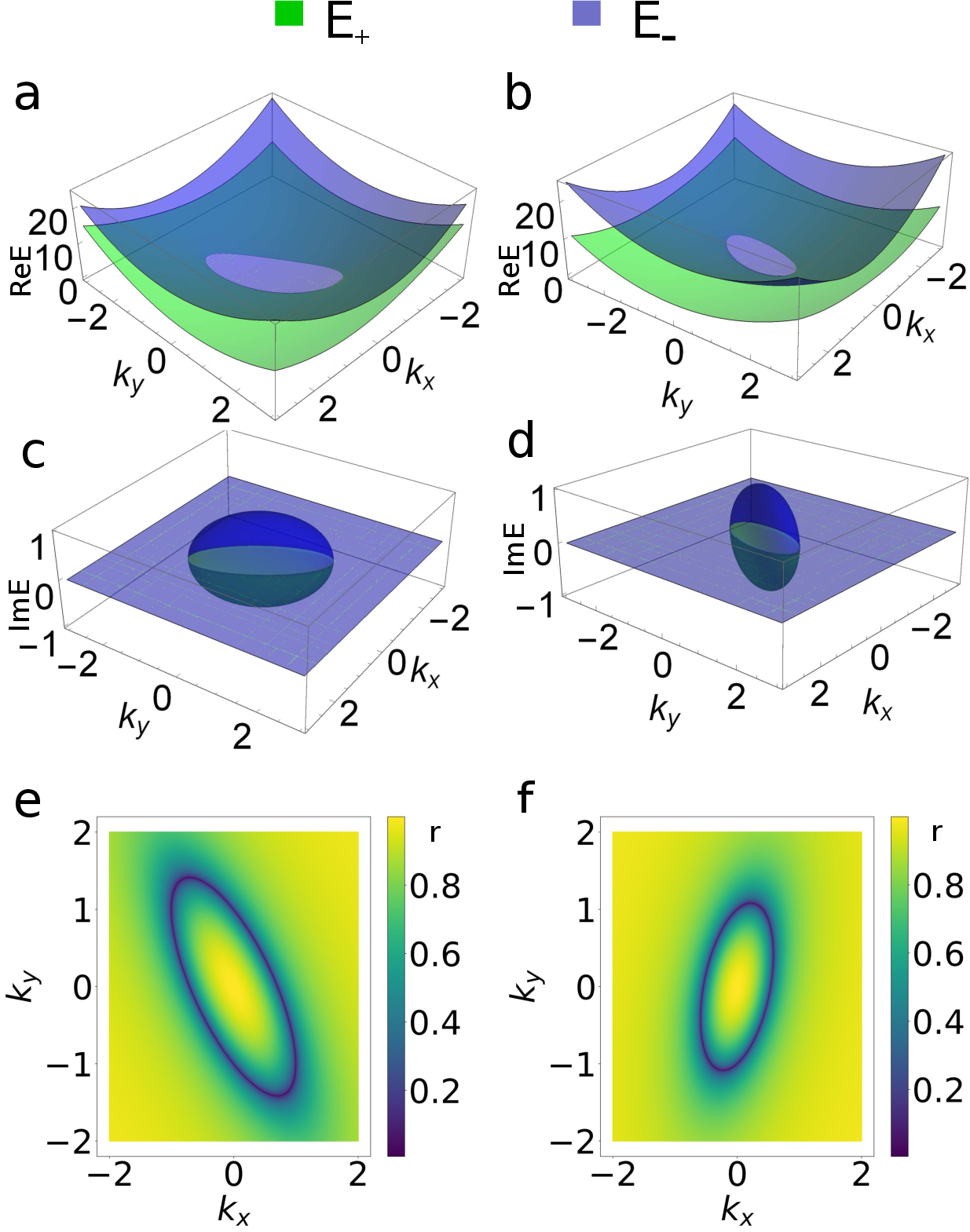}
\caption{\textbf{Exceptional rings in $p-$wave magnet-FM junctions.} (a), (b) Real and (c), (d) imaginary parts of the eigenvalues for \textcolor{black}{$\phi=\pi$ and $\phi=\pi/4$} 
respectively. Note that eigenvalues merge along the ellipses. The phase rigidity, $r$, is plotted in (e) and (f). Note that $r$ goes to zero along the ellipses, signaling their exceptional nature. We find that the orientation and size of the exceptional ring depends on $\phi$, i.e, on the orientation of the N\'eel vector. Here we choose $t=1$, $\lambda=1$ \textcolor{black}{$\theta=\pi/2$, $J=1$} and $\gamma=\textcolor{black}{-}1$.}
\label{Fig_ring}
\end{figure}

This NH Hamiltonian can be expressed in the form $\Tilde{H} = \epsilon_0 + \mathbf{d} \cdot \boldsymbol{\sigma}$, where \textcolor{black}{$\epsilon_0 = t(k_x^2 + k_y^2) - i\Gamma$}, $\epsilon_0 \in \mathbb{C}$ and the complex vector $\mathbf{d} = \mathbf{d}_R + i\mathbf{d}_I$, with $\mathbf{d}_R, \mathbf{d}_I \in \mathbb{R}^3$. For our specific model, the real part is given by $\mathbf{d}_R = (-\lambda k_y + Jk_x \sin\theta \cos\phi,\ \lambda k_x + Jk_x \sin\theta \sin\phi,\ Jk_x \cos\theta)$, and the imaginary part is $\mathbf{d}_I = (0,\ 0,\ -\gamma)$. The energy eigenvalues are \textcolor{black}{$E_{\pm} = \epsilon_0 \pm \sqrt{\mathbf{d}_R^2 - \mathbf{d}_I^2 + 2i \mathbf{d}_R \cdot \mathbf{d}_I}$}. NH degeneracies occur when the conditions $\mathbf{d}_R^2 = \mathbf{d}_I^2$ and $\mathbf{d}_R \cdot \mathbf{d}_I = 0$ are simultaneously satisfied. Applying these to the $p$-wave-FM junction leads to the following constraints,

\begin{equation}
\textcolor{black}{\gamma^2 =} 
\begin{aligned}[t]
&\textcolor{black}{\big(-\lambda k_y + Jk_x \sin\theta \cos\phi\big)^2} \\
&\textcolor{black}{+ \big(\lambda k_x + Jk_x \sin\theta \sin\phi\big)^2} \\
&\textcolor{black}{+ \big(Jk_x \cos\theta\big)^2}
\end{aligned}
\label{eq_EP_rules_a}
\end{equation}

\begin{equation}
\textcolor{black}{\gamma\, J\, k_x \cos\theta = 0}
\label{eq_EP_rules_b}
\end{equation}

We discard the trivial case with $\gamma=0$. From Eq.~\ref{eq_EP_rules_b}, we find that the NH degeneracies occur for $\theta=\pi/2$. This implies that the Néel vector of the $p$-wave magnet should lie in the $x–y$ plane, utilizing the non-collinear coplanar spin arrangement. From the second condition, we obtain $Jk_x\cos\theta=0$, which makes the last term of the first condition vanish. As a result, the exceptional degeneracy conditions turn into a generalized equation of a conic section,

\textcolor{black}{
\begin{equation}
\begin{aligned}
 k_x^2(\lambda^2+2\lambda J\sin\phi + J^2)\\
 -2\lambda Jk_xk_y\cos\phi + \lambda^2k_y^2 - \gamma^2=0.
\end{aligned}
\label{Eq_conic}
\end{equation}}

The discriminant $X$ is given by~\cite{sharma2005text}, 
\textcolor{black}{
\begin{equation}
    X= 4\lambda^2J^2\cos^2\phi-4\lambda^2(\lambda^2+2\lambda J\sin\phi+J^2).
    \label{Eq_X}
\end{equation}}

Going forward, we fix the parameters to $t = 1$, $\gamma = \textcolor{black}{-}1$, \textcolor{black}{$J=1$} and $\lambda = 1$ for simplicity, \textcolor{black}{unless stated otherwise.}. \textcolor{black}{The parameter $\Gamma$ appears only in the expression for $\epsilon_0$ as an energy shift, and it uniformly shifts the eigenvalues without affecting their structure or physical implications. Therefore, for simplicity, we have fixed $\Gamma = 0$ in all our calculations.} We next present the various types of NH degeneracies that can be designed in this junction. \\

\noindent \textit{\textbf{Symmetry protected exceptional rings--}} For $\theta = \pi/2$ and general values of $\phi$ and $J$, the discriminant $X$ becomes negative, resulting in an elliptical shaped exceptional ring. For this condition, the general equation of a conic section turns into

\begin{equation}
 k_x^2(\lambda^2+2\lambda J\sin\phi + J^2)-2\lambda Jk_xk_y\cos\phi +\lambda^2k_y^2 - \gamma^2=0.
\label{Eq_conic2}
\end{equation}

This can be rewritten as, $A k_x^2 + B k_xk_y + C k_y^2 - \gamma^2=0$, with $A=(\lambda^2+2\lambda J\sin\phi + J^2)$, $B=-2\lambda Jk_xk_y\cos\phi$ and $C=\lambda^2$. Applying the following coordinate transformation $k_x= K_X\cos\eta -K_Y\sin\eta,  k_y= K_X\sin\eta + K_Y\cos\eta$, transforms the equation to $A'K_X^2+C'K_Y^2=\gamma^2$, which is the equation of an ellipse. Here $\eta$ is given by $\tan2\eta= \frac{B}{A-C}$ and $ A'=A\cos^2\eta+ B\cos\eta\sin\eta + C\sin^2\eta$, $C'=A\sin^2\eta - B\cos\eta\sin\eta + C\cos^2\eta$. We note that the lengths of the semi-major and semi-minor axes depend on the magnet and junction parameters $(\lambda, \phi, J, \gamma)$, thereby enabling control over the ER.

We present the real and imaginary parts of the eigenvalues in Fig.~\ref{Fig_ring} for two sets of parameter values \textcolor{black}{$\phi=\pi$ and $\phi=\frac{\pi}{4}$}, respectively. The eigenvalues merge along the elliptically shaped rings. We also note that the orientation and the size of the ER can be directly controlled by changing the N\'eel vector of the $p$-wave magnet. 

We further verify the coalescence of eigenvectors along the rings by calculating the phase rigidity~\cite{heiss2012physics}, 

\begin{equation}
r = \frac{\langle \Psi_L | \Psi_R\rangle}{ \langle \Psi_R | \Psi_R \rangle}.
\end{equation}
Here $ \Psi_L$ and $ \Psi_R$ denote the left and right eigenvectors, respectively.
\textcolor{black}{In a non-Hermitian Hamiltonian, the Hilbert space is spanned by both the right and left eigenstates ~\cite{brody2013biorthogonal}, which are distinct in contrast to the Hermitian case. The orthogonality between left and right eigenstates is established through \emph{biorthogonalization}, defined as $\langle \Psi_L^{m} | \Psi_R^{n} \rangle = \delta_{mn}$. At EPs, the left and right eigenstates become orthogonal to each other, reflecting the coalescence of eigenvalues and eigenvectors.}
Due to bi-orthogonalization, $r\rightarrow 0$ near an exceptional degeneracy and approaches unity away from them. \textcolor{black}{In the limit \( r \to 0 \), i.e., in the vicinity EPs, the right and left eigenstates become orthogonal, resulting in the coalescence of the associated physical states ~\cite{banerjee2023non}. From an experimental standpoint—especially in sensing applications—operating near \( r \to 0 \) can greatly enhance sensitivity, as even small perturbations can lead to substantial and detectable changes in the signal due to the non-Hermitian degeneracy.} We have confirmed the exceptional nature of the tunable ring by examining the phase rigidity, as shown in Fig.~\ref{Fig_ring}(e)-(f). We find that the rigidity vanishes to zero along the ring, as expected, while becoming finite away from it.

We emphasize that the ERs generated in our $p$-wave magnet-FM junctions are protected by chiral symmetry. In the present case, $\sigma_z$ serves as the chiral symmetry operator. The effective Hamiltonian $\Tilde{H}$ satisfies the chiral symmetry condition, i.e., $\Tilde{H} = -\sigma_z^{\dagger} \Tilde{H}^{\dagger} \sigma_z$, provided that the component of $\mathbf{n}$  along \textcolor{black}{$z$} vanishes in the absence of a constant energy shift. This condition holds specifically for $\theta =\pi/2$, establishing that the ER is indeed symmetry-protected. \textcolor{black}{The ER is protected by chiral symmetry and remains robust under generic disorder or perturbations that preserve this symmetry. In particular, when the Néel vector lies in the $x-y$ plane, i.e., $\theta = \pi/2$, variations in the azimuthal angle $\phi$ or other system parameters do not break the chiral symmetry, and thus the ER persists. However, if the disorder or perturbation introduces a term proportional to $\sigma_z$ in the Hamiltonian, the chiral symmetry is explicitly broken, and the protection of the ER is lost.} Next, we discuss the other regime of the junction parameter space for which parallel exceptional lines appear at the interface. \\

\begin{figure}
\centering
\includegraphics[width=0.48\textwidth]{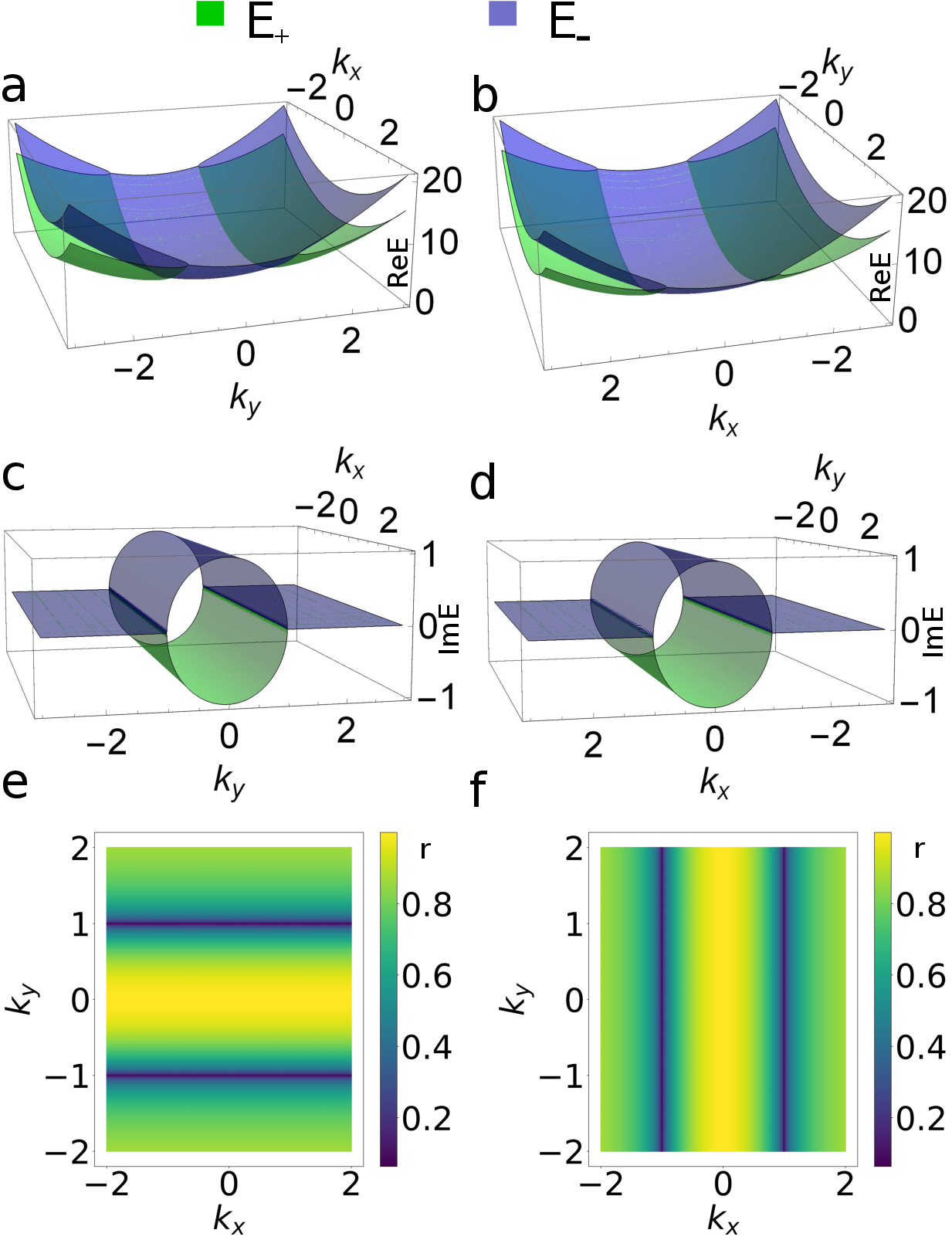}
\caption{\textbf{Exceptional lines in $p-$wave magnet-FM junctions.} (a), (b) Real and (c), (d) imaginary parts of the eigenvalues for \textcolor{black}{$(\phi,\lambda)=(3\pi/2, 1)$ and $(\phi,\lambda)=(3\pi/2, 0)$}, respectively. Note that eigenvalues merge along a pair of parallel lines. The phase rigidity, $r$, is plotted for the above conditions in (e) and (f). The deep blue color along two parallel lines indicates $r=0$, which confirms the coalescing of eigenvectors. We observe that the orientation and distance between lines depend on junction parameters and the orientation of the N\'eel vector. Here we choose $t=1$, \textcolor{black}{$J=1$, $\theta=\pi/2$} and $\gamma=\textcolor{black}{-}1$.} 
\label{Fig_lines}
\end{figure}


\noindent\textit{\textbf{Parallel exceptional lines--}} Let us now consider the orientation of the N\'eel vector which yields exceptional lines rather than rings. There are two qualitatively different cases, one with a finite Rashba strength $\lambda$ and the other with vanishing $\lambda$.

Consider first the case with finite $\lambda$ and choosing \textcolor{black}{$(\phi, \lambda) = \left( \frac{3\pi}{2},  1\right)$}. In this situation we find, from Eq.~\ref{Eq_conic}, that the discriminant $X$ vanishes, leading to $k_y = \pm \frac{\gamma}{\lambda}$. This corresponds to two parallel exceptional lines along the $k_x$-axis as shown in Fig.~\ref{Fig_lines}(a)-(c). Remarkably, the separation between the two exceptional lines can be tuned and is determined by the Rashba coupling strength $\lambda$ and the coupling between the FM and $p$-wave magnet, $\gamma$.

In the second case, when the Rashba term vanishes, the discriminant $X$ again goes to zero. Choosing \textcolor{black}{$(\phi,\lambda) = \left( \frac{3\pi}{2},  0\right)$}, we find that the condition for exceptional behavior becomes $k_x = \pm \frac{\gamma}{J}$. This results in two exceptional parallel lines, which now lie along the $k_y$-axis as presented in Fig.~\ref{Fig_lines}(b)-(d). The separation is again governed by the junction coupling parameter $\gamma$ and $p$-wave magnet strength $J$.

We confirm the exceptional nature of parallel degeneracies of both the above cases by computing the phase rigidity $r$. We find that $r$ vanishes along the two sets of parallel lines as shown in Fig.~\ref{Fig_lines}(e)-(f). This confirms that not only the eigenvalues merge but also the eigenstates coalesce. In both cases, the vanishing discriminant leads to the emergence of two parallel exceptional lines. Most notably, their orientation and separation are controlled by junction and magnet parameters $\gamma$, $\lambda$, and $J$. \textcolor{black}{In contrast to ERs, the exceptional lines emerge only for specific fine-tuned values of $\phi$, and therefore, EP lines are not symmetry protected.}
We note that the emergence of exceptional lines in the latter case differs from the emergence of EPs in the $d$-wave altermagnet-FM junction, where Rashba interaction was essential~\cite{reja2024emergence}. Furthermore, it can be shown that for $k_x = 0$ and $\theta \neq \pm \pi/2$, the junction exhibits a pair of EPs~\cite{yoshida2019symmetry,bergholtz2021exceptional,wang2019non} -- this corresponds to the $p$-wave term vanishing. \textcolor{black}{Replacing the $p-$wave magnet with an $f-$wave magnet in this setup, we uncover intriguing results, namely intricately-shaped exceptional contours and EPs. For more details, see the appendix.}\\


\noindent \textit{\textbf{Summary and outlook--}} We have proposed $p$-wave magnet–FM junctions as a versatile platform to explore non-Hermitian physics. We demonstrated that the junction hosts exceptional contours -- lines and rings -- which can be directly controlled by the orientation of the $p$-wave magnet N\'eel vector. We found that the separation between the parallel exceptional lines and the size of the exceptional rings depend on the Néel vector, the coupling at the junction, and the strength of the Rashba interaction. \textcolor{black}{While predicted $p$-wave magnets are rare, CeNiAsO is a promising candidate that has recently been studied ~\cite{hellenes2023p, chakraborty2024highly}. There are no special requirements on the ferromagnet, other than a good lattice match with the $p$-wave magnet. As such conventional ferromagnets could be used for our proposal. In addition to the formation of the interface, another important aspect is to probe the existence of the exceptional contours. In this direction, we note that our phase rigidity predictions should be testable, as has been previously measured in several other systems~\cite{ding2016emergence}. Furthermore, the non-Hermitian spectra in altermagnet-ferromagnet junctions may be directly amenable to surface spectroscopy measurements, such as angle resolved photoemission spectroscopy.}
\textcolor{black}{As noted in Ref.~\cite{hellenes2023p}, the relevant energy scale for the considered $p$-wave magnetic material CeNiAsO is approximately 7.6~K. This low-energy scale indeed poses experimental challenges to maintaining the $p$-wave magnetism. Nevertheless, we are hopeful that this can still allow an observation of our predicted exceptional physics.} In closing, we note that exceptional rings have so far been observed~\cite{cerjan2019experimental,zhen2015spawning,xu2022observation} or proposed~\cite{liu2021higher,xu2017weyl} in only a very limited number of systems. Our work introduces a novel platform for exploring non-Hermitian physics and controlling exceptional degeneracies, while highlighting an as-yet-unexplored aspect of odd-parity magnets. \\


\noindent \textit{\textbf{Acknowledgments--}} We thank A. Banerjee and A. Bose for useful discussions. M.A.R. is supported by a graduate fellowship of the Indian Institute of Science. A.N. acknowledges the DST MATRICS grant (MTR/2023/000021).

\bibliography{ref.bib}

\vspace{0.5cm}

\renewcommand{\theequation}{A\arabic{equation}}
\renewcommand{\thesection}{A\arabic{section}}
\renewcommand {\thefigure}{A\arabic{figure}}
\setcounter{equation}{0}
\setcounter{figure}{0}
\noindent \textit{\textbf{Appendix: Exceptional contours and EPs in $f-$ wave magnet ferromagnet junctions--}}

\begin{figure}
\centering
\includegraphics[width=0.48\textwidth]{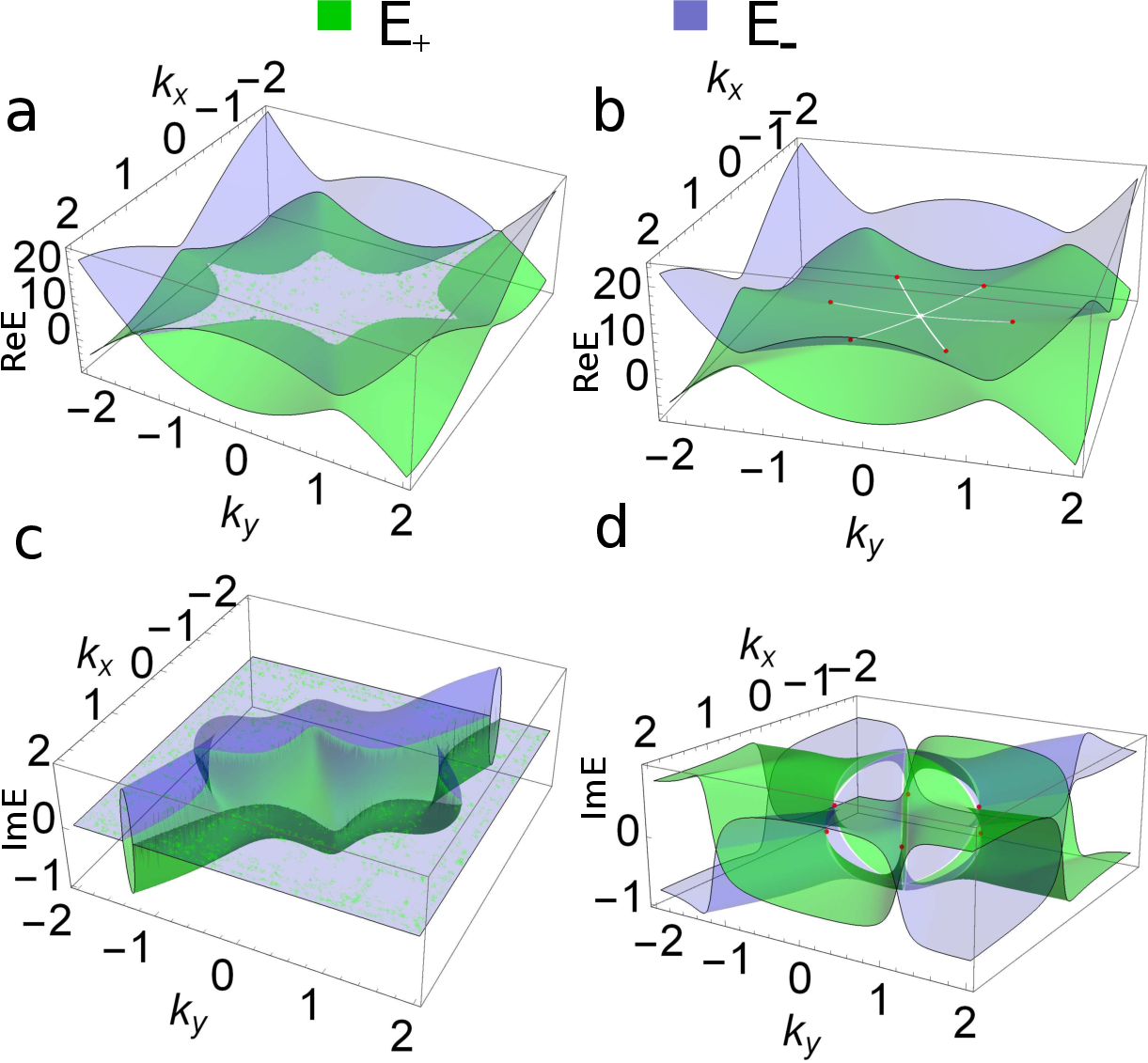}
\caption{\textbf{Exceptional contours and EPs in $f$-wave magnet-FM junctions.} (a)-(b) Real and (c)-(d) imaginary parts of the eigenvalues for $\theta = \pi/2$ and $\theta = \pi/4$, respectively. Note that eigenvalues merge along an intricate structure of contours for $\theta = \pi/2$. On the other hand, they merge at specific points for $\theta \neq \pi/2$ case. The red dots represent the locations of the EPs. Here we choose $t=1$, $\lambda=1$, $J=1$, $\phi=\pi/4$, $\gamma=-1$ and  $\Gamma=0$.}
\label{Fig_fwave}
\end{figure}

We consider an $f-$wave magnet-FM junction described by the following Hamiltonian~\cite{fukaya2025superconducting},

\begin{equation}
\begin{aligned}
\Tilde{H} 
&= t\,(k_x^2 + k_y^2)\,\sigma_0 
+ \lambda\,(-k_y \sigma_x + k_x \sigma_y) \\
&\quad + J (\mathbf{n}\cdot \boldsymbol{\sigma}) [ k_x(k_x^2 - 3k_y^2) \cos 3\beta 
+ k_y(3k_x^2 - k_y^2) \sin 3\beta \big] \\
&\quad+ \Sigma_L .
\end{aligned}
\label{eq_f_ham}
\end{equation}

Here $\beta$ defines the orientation of Fermi surface of the $f$-wave magnet. For simplicity, we set $\beta=0$. Then, the conditions for the occurrence of non-Hermitian degeneracy turns out to be

\begin{equation}
\begin{aligned}
\gamma^2 &= \big(-\lambda k_y 
+ Jk_x(k_x^2 - 3k_y^2)\sin\theta\cos\phi\big)^2 \\
&\quad + \big(\lambda k_x 
+ Jk_x(k_x^2 - 3k_y^2)\sin\theta\sin\phi\big)^2 \\
&\quad + \big(Jk_x(k_x^2 - 3k_y^2)\cos\theta\big)^2,
\end{aligned}
\label{eq_f_EP_rules_a}
\end{equation}

\begin{equation}
\gamma\, J\, k_x(k_x^2 - 3k_y^2)\cos\theta = 0.
\label{eq_f_EP_rules_b}
\end{equation}

The condition in Eq.~\ref{eq_f_EP_rules_b} can be satisfied in three distinct ways: 
$\theta = \pi/2$, or $k_x = 0$, or $k_x^2 = 3k_y^2$, as well as by combinations of these conditions. We examine these conditions by categorizing them into two broad cases, discussed in detail below. \\

\noindent\textbf{Case 1:} For $\theta = \pi/2$, the system exhibits exceptional contours with an intricate structure following the equation $\gamma^2= (-\lambda k_y 
+ Jk_x(k_x^2 - 3k_y^2)\sin\cos\phi)^2 + (\lambda k_x 
+ Jk_x(k_x^2 - 3k_y^2)\sin\phi)^2$. This follows from the Eq.~\ref{eq_f_EP_rules_a} with $\theta = \pi/2$. The merging of the real and imaginary eigenvalues along such a contour is shown in Fig.~\ref{Fig_fwave}(a) and (c). \\

\noindent\textbf{Case 2:} For $\theta \neq \pi/2$, six EPs appear at the coordinates 
$(0, \pm\frac{\gamma}{\lambda})$ and $(\pm\frac{\sqrt{3}\,\gamma}{2\lambda}, \pm\frac{\gamma}{2\lambda})$. The red dots in Fig.~\ref{Fig_fwave}(b) and (d) denote the EPs for this case. For a general value of $\beta$, the locations of these EPs shift accordingly, but their number and qualitative features remain unchanged.

\end{document}